# Synthesis of stable cerium oxide nanoparticles coated with phosphonic acid-based functional polymers

**Ameni Dhouib[1,†], Braham Mezghrani[1,†], Giusy Finocchiaro[1,2], Rémi Le Borgne[3], Mathéo Berthet[4], Bénédicte Daydé-Cazals[4], Alain Graillot[4], Xiaohui Ju[5,6*], Jean-François Berret[1]***

*[1]Université Paris Cité, CNRS, Matière et systèmes complexes, 75013 Paris, France*
*[2]Institute of Photonics and Electronics of the Czech Academy of Sciences, Chaberská1014/57, 182 51 Prague, Czech Republic*
*[3]Université Paris Cité, CNRS, Institut Jacques Monod, F-75013 Paris, France*
*[4]Specific Polymers, ZAC Via Domitia, 150 Avenue des Cocardières, 34160 Castries, France*
*[5] Department of Chemistry and Biochemistry, Mendel University in Brno, Zemedelska 1, 613 00 Brno, Czech Republic*
*[6] Department of Surface and Plasma Science, Faculty of Mathematics and Physics, Charles University, Prague, 181 00, Czech Republic*

**Abstract**: Functional polymers, such as poly(ethylene glycol) (PEG) terminated with a single phosphonic acid, hereafter PEG$_{ik}$-Ph are often applied to coat metal oxide surfaces during post synthesis steps, but are not sufficient to stabilize sub-10 nm particles in protein-rich biofluids. The instability is attributed to the weak binding affinity of post-grafted phosphonic acid groups, resulting in a gradual detachment of the polymers the surface. Here, we assess these polymers as coating agents using an alternative route, namely the one-step wet-chemical synthesis, where PEG$_{ik}$-Ph is introduced with cerium precursors during the synthesis. Characterization of the coated cerium oxide nanoparticles indicates a core-shell structure, where the cores are 3 nm cerium oxide and the shell consists of functionalized PEG polymers in a brush configuration. Results show that cerium oxide nanoparticles coated with PEG$_{1k}$-Ph and PEG$_{2k}$-Ph are of potential interest for applications as nanomedicine due to their high Ce(III) content and increased colloidal stability in cell culture media. We further demonstrate that the cerium oxide nanoparticles in the presence of hydrogen peroxide show an additional absorbance band in the UV-vis spectrum, which is attributed to Ce-O$_2{}^{2-}$ peroxo-complexes and could be used in the evaluation of their catalytic activity for scavenging reactive oxygen species.

## I – Introduction

In recent years, nanostructured cerium oxide (CeO$_2$) is receiving much attention as a potential antioxidant agent for biomedical applications thanks to its enzyme mimetic catalytic properties.[1-3] The ability of cerium oxide nanoparticles (CNPs) to act as an antioxidant is derived from their quick and expedient mutation of the oxidation state between Ce(IV) and Ce(III) by adjusting their electronic configuration and exhibiting oxygen vacancies, or defects, in the lattice structure.[2,4-7]

The unique oxygen buffering behavior of cerium oxide has been studied extensively in the surface science and catalysis field since cerium oxide can serve as an ideal industrial catalyst.[8-9] While the biomedical antioxidant properties of nanoscale $CeO_2$ are promising, there is a great challenge to apply existing CNPs for clinical use.[3,10-11] For medical applications, CNPs must be functionalized by proper hydrophilic coatings to avoid aggregation/accumulation in the human cellular environment and to enhance their biocompatibility, while retaining their antioxidant properties.[1,12] Extensive research has shown that CNP biomimetic activity was greatly influenced by its electrostatic surface charges and coating adlayer, due to the interference with electron transfer between surface Ce(III) and solution oxidants.[13-14] Several parameters are critical regarding hydrophilic polymer coatings, such as the nature of the binding moieties to the surface, the polymer density, polymer charges, and the coating thickness.[12,15-18] The fact that CNPs with the smallest hydrodynamic diameter and thinnest coating display the fastest kinetics to the reaction between $H_2O_2$ and Ce(III) suggests that the adlayer thickness plays a key role in the rate of oxidation of the substrate.[14]

A wide range of techniques has been applied for the synthesis and stabilization of CNPs through polymer coatings.[10-11] Wet chemical precipitation synthesis is one of the most commonly used methods to build up nanoparticle crystals from molecular precursors, such as cerium nitrate hexahydrate. Hydrophilic polymer coatings can be functionalized to the CNPs surfaces through two different approaches. It can be either achieved through one-step wet-chemical synthesis, where the polymers are introduced together with the cerium precursors during synthesis.[1,16] These polymers further act as capping agents limiting the nanocrystal growth and preventing their agglomeration. Another effective route of polymer coatings onto CNPs is the two-step post-synthesis process, where the polymers are grafted to CNPs surfaces after $CeO_2$ nanoparticles synthesis.[15,19] This approach is achieved through grafting-from or co-assembly grafting techniques, as described in our previous work.[15,20-22] The latter strategy involves the spontaneous adsorption of polymers through electrostatics[15,20,23-25] and the phase transfer of coating agents between organic and aqueous solvents.[14,26]

Many polymers have been proven suitable for CNP hydrophilic coating for biomedical applications due to their superior colloidal stability and biocompatibility, such as dextran,[12-13] chitosan, poly(acrylic acid),[16,27-28] oleic acid,[14,29] and poly (ethylene glycol) (PEG).[15,26,30-31] PEG is among the most widely used and mature polymers developed for such purposes.[31] Approved by regulatory and control agencies, PEG offers also substantial advantages. It is a flexible macromolecule that can be synthesized with narrow molar mass dispersity. Moreover, PEG was also found to follow accurate polymer dynamics predictions, allowing quantitative evaluation of the chain conformation at solid-liquid interfaces.[32-33]

In our previous work, we developed a platform for the coating of metal oxide particles (e.g. $CeO_2$, $\gamma$-$Fe_2O_3$, $Al_2O_3$ and $TiO_2$)[20] based on PEG functional polymers.[15,20-21] The platform was designed to synthesize (co)polymers carrying multiple functions along their backbone: i) phosphonic acid ($R$-$H_2PO_3$) as anchoring groups to the metal surface,[34-35] ii) PEG pendant chain for colloidal stability and protein resistance[32] and iii) amine terminated PEGylated chain for further covalent functionalization.[36] In controlled physicochemical conditions, phosphonic acid PEG copolymers spontaneously adsorb to the metal oxide surfaces via the phosphonic acid groups, with the PEG forming a dense brush of extended chains. The robust non-covalent anchoring to the surface was ascribed to the presence of the three oxygens of the phosphonic acid group, allowing for mono, bi-, and tridentate binding modes to cerium atoms.[34,37-38] In vitro mammalian cell assays[21-22,36] and in vivo biodistribution experiments in mice[21,36,39] have shown that phosphonic acid PEG copolymers are effective coating agents for iron and cerium oxide nanoparticles, as they significantly reduce cellular uptake and extend the lifetime of particles in the blood circulation for several hours. While multi-phosphonic acid PEG copolymers have clear advantages over monophosphonic acid PEG in terms of colloidal stability,[20] they do exhibit some drawbacks. First, their synthesis is more challenging than the monophosphonic acid PEG. Second, these polymers are statistical, with a relatively broad molar-mass dispersity. For such samples, the diverse mixture of short and long chains may behave differently compared to the average value.

In this work, we investigate monophosphonic acid PEG polymers of molecular weight 1, 2, and 5 kDa as CNPs hydrophilic polymer coating, in combination with the one-pot synthesis and functionalization method. To our knowledge, this is the first time that monophosphonic acid PEG functional polymers are used for this purpose in the context of CNP synthesis.[11] Previous attempts to implement PEG homopolymers as coating failed to reveal the chemical coupling between ethylene oxide monomers and $CeO_2$ surface.[25,30,40-41] The objective is to take advantage of the high affinity of the phosphonic acid groups for cerium to strengthen the anchoring of polymers terminated with a single phosphonic acid. In this work we show that CNPs synthesized in the presence of monophosphonic acid PEG polymer for controlling for nanocrystal growth generate 3 nm $CeO_2$ cores presenting a high fraction of Ce(III) on the surface, and combining remarkable stability in the culture media. We have also developed a spectrophotometric approach that reveals an additional band at 365 nm in the CNP absorption spectrum in the presence of hydrogen peroxide ($H_2O_2$). This band is attributed to Ce-$O_2^{2-}$ peroxo-complexes[42-47] and its amplitude could be potentially used in the evaluation of catalytic properties of cerium oxide nanoparticles as enzyme mimetics.

## II - Materials and Methods

### II.1 – Materials

Cerium oxide nanoparticles with a diameter of 7.8 nm (CNP8) were synthesized by thermo-hydrolysis of Ce(III) nitrate hexahydrate and kindly provided to us by Solvay (Centre de Recherche d'Aubervilliers, France) as a 250 g $L^{-1}$ aqueous dispersion (pH 1.5).[15] A transmission electron microscopy (TEM) image of the CNP8 particles along with the size distribution is provided in Supporting Information S1. The PEG polymers functionalized by one or several phosphonic acid moieties were synthesized by SPECIFIC POLYMERS®, France (http://www.specificpolymers.fr/). These are 3 linear polymers synthesized by functionalization of PEG$_{ik}$ (i = 1, 2 or 5) precursors with a single phosphonic acid group as end-group, and a statistical copolymer with a methacrylate backbone and pending PEG$_{2k}$ chains and 5 phosphonic acid groups. For the abbreviations of the polymers, we refer to the terminology published in ref.[21]: phosphonic acid terminated PEG polymer are termed PEG$_{1k}$-Ph, PEG$_{2k}$-Ph and PEG$_{5k}$-Ph respectively (Fig. 1a), whereas the copolymer reads MPEG$_{2k}$-*co*-MPh (Fig. 1b). For the latter, MPEG$_{2k}$ denotes a PEG methacrylate monomer, while MPh represents a methacrylate monomer bearing a phosphonic acid functional group. For the PEG$_{ik}$-Ph polymers, size exclusion chromatography (SEC) was used to determine the number and weight-averaged molecular weights $M_n$ and $M_w$, as well the molar mass dispersity Đ (Supporting Information S2). The molar composition of the copolymer MPEG$_{2k}$-*co*-MPh was determined by $^1$H NMR integrated intensities obtained on the phosphonate ester precursor, considering NMR signals from the methyl ester associated to the MPh group and the ethylene oxide groups of the PEGylated chains. From the ratio between these intensities, we obtained a copolymer with 6.2 PEGylated chains and 5.4 phosphonic acid groups in average, leading to the formula (MPEG$_{2k}$)$_6$-*co*-MPh$_5$. The presence of the phosphonic acid unit was confirmed by $^{31}$P NMR with a signal between 16 and 21 ppm (Supporting Information S3). For the copolymer, $M_w$ was obtained from static light scattering and the Zimm representation of the scattered intensity versus concentration,[20] whereas a molar-mass dispersity of 1.8 was obtained by SEC on PolyPore column using THF as eluent and polystyrene standards. Synthesis details and characterization can be found in previous reports.[21-22,39] Dulbecco's Modified Eagle Medium (DMEM), foetal bovine serum (FBS) and penicillin–streptomycin were purchased from Gibco, Life Technologies. Water was deionized with a Millipore Milli-Q Water system. All products were used without further purification. HNO$_3$ and NH$_4$OH were used to adjust the pH.

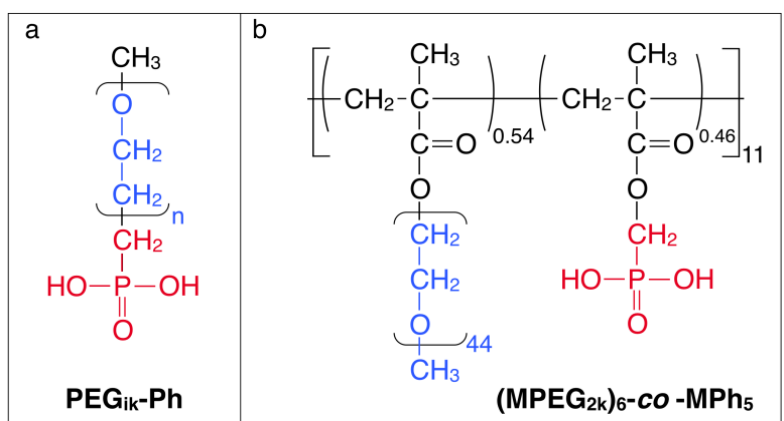

***Figure 1***: *Chemical structures and abbreviations of the polymers used in this work.* ***(a)*** *Poly(ethylene glycol) with the molecular weight i kDa (i = 1, 2, and 5) terminated with a single phosphonic acid and abbreviated as PEG$_{ik}$-Ph.[21]* ***(b)*** *Statistical copolymer [(MPEG$_{2k}$)$_{0.54}$-co-MPh$_{0.46}$]$_{11}$ made from MPEG$_{2k}$ and MPh monomers, wherein MPEG$_{2k}$ refers to a PEG methacrylate macromonomer with PEG molecular weight of 2 kDa and MPh to a methacrylate monomer bearing a phosphonic acid functional group. The molar ratio of each comonomer was determined from $^1$H NMR (Supporting Information S3). For simplicity, the copolymer will be denoted hereafter as (MPEG$_{2k}$)$_6$-co-MPh$_5$.*

| Polymers | $M_n$ (Da) | $M_w$ (Da) | Molar mass dispersity Đ | Phosphonic acid/polymer |
|---|---|---|---|---|
| PEG$_{1k}$-Ph | 1060 | 1320 | 1.24 | 1.0 |
| PEG$_{2k}$-Ph | 1670 | 1750 | 1.05 | 1.0 |
| PEG$_{5k}$-Ph | 4190 | 4400 | 1.05 | 1.0 |
| (MPEG$_{2k}$)$_6$-*co*-MPh$_5$ | 11300 | 20300 | 1.8 | 5.4 |

***Table I*** : *Molecular characteristics of the phosphonic acid PEGylated polymers and copolymer synthesized in this work. The molecular weight and molar mass dispersity Đ of the PEG$_{ik}$-Ph precursors were determined from size exclusion chromatography (Supporting Information S2).*

The average numbers of phosphonic acids and PEG segments in $(MPEG_{2k})_6$-co-$MPh_5$ were estimated from the molar ratio and the weight-average molecular weight $M_w$.[20]

## II.2 – Synthesis of cerium oxide nanoparticles

The CNP dispersions were prepared by a wet-chemical one-pot method. Briefly, 35 µmol of phosphonic acid PEG polymers were dissolved in 250 µL of DI-water at pH 1.4. Similarly, 35 µmol of cerium nitrate hexahydrate (Ce(NO$_3$)$_3$·6H$_2$O) were dissolved in 250 µL of DI-water at pH 1.4, then added and mixed thoroughly with the polymer solution for one hour at room temperature. In this first series of experiments, the Ce-to-phosphonic-acid stoichiometry was set to 1. Then, 50 µL of 30% ammonium hydroxide was added to the previous solution and the CNP nucleation and growth were allowed to proceed for 24h at 60 °C. The pH was subsequently adjusted to 7.4. The CNP dispersion was ultra-filtrated using an Amicon cell with a 100 kDa molecular weight cut-off membrane (Millipore Inc.) and centrifuged at 3000 rpm for 15 minutes to remove the excess polymers. In the Ce(III) ions oxidation reaction leading to the CNP formation, the PEG-based polymers are allowed to slow down the nanocrystal growth and at the same time stabilize the particles by steric repulsion. In a second series, the scaling up of a chemical reaction was performed by increasing the individual components by a factor of 10 (350 µmol of (Ce(NO$_3$)$_3$·6H$_2$O) and of phosphonic acid PEG polymers in a total of 3 mL). To finally investigate the role of concentration and stoichiometry, the synthesis was performed by changing the Ce-to-phosphonic-acid stoichiometry (from 1:1 to 1:0.5 and 1:0.1), and reducing the molar contents of the constituents by factors 2, 4 and 8, while keeping the synthesis volumes unchanged.

## II.3 – Small- and Wide-Angle X-Ray Scattering (SAXS,WAXS)

X-ray scattering was carried out using an Empyrean (PANALYTICAL) diffractometer equipped with a multichannel PIXcel 3D detector and a Cu Kα X-ray source (1.54187 Å). Samples were deposited on a monocrystalline Si substrate, with a spinner movement (rotation time 1 s). A 1/16° divergence slit, a 1/8° anti-scatter slit and a 10 mm mask were installed before the samples. Typically, each pattern was recorded in the $\theta - \theta$ Bragg-Brentano geometry in the 10°-20° $2\theta$ range (0.0263° for 600 s), where $2\theta$ denotes the scattering angle.

## II.4 – Dynamic light scattering and zeta potential

The scattering intensity $I_S$ and the hydrodynamic diameter $D_H$ was determined on CNP dispersions using a NanoZS Zetasizer (Malvern Panalytical) at the concentration $c_{CNP}$ = 0.1 − 20 g L$^{-1}$ (T = 25 °C). The scattering intensity as a function of concentration showed that in the dilute regime $I_S$ is proportional to $c_{CNP}$. In the dynamic mode, the second-order autocorrelation function was recorded in triplicate and analyzed using the cumulant algorithms to determine the average diffusion coefficient and the intensity distribution of the hydrodynamic diameter. The $D_H$ was calculated according to the Stokes-Einstein relation $D_H = k_B T / 3\pi\eta D_0$ where $k_B$ is the Boltzmann constant, $T$ is the temperature, and $\eta$ is the solvent viscosity (0.89 mPa s). Laser Doppler velocimetry using the phase analysis light scattering mode (NanoZS Zetasizer) was performed to determine the electrophoretic mobility and zeta potential. For colloidal stability studies, a few microliters of a concentrated CNP dispersion were poured and homogenized in 1 mL of the cell culture medium, and the scattering intensity $I_S(t)$ and hydrodynamic diameter $D_H(t)$ were simultaneously measured over time for a period of one month (T = 25 °C). In this protocol, we also verify that the scattering intensity coming from the CNPs was much higher than that of the proteins in the culture medium and that the measured $D_H(t)$ were those of the nanoparticles only.

## II.5 – UV-Visible spectrophotometry

The absorbance of CNP aqueous dispersions was measured with a UV-visible spectrophotometer (JASCO, V-630) equipped with a temperature controller. The CNP absorption spectra $A_{CNP}(\lambda)$ were recorded in the range $\lambda = 190 - 800$ nm at room temperature (T = 25 °C). According to the Beer-Lambert law, $A_{CNP}(\lambda)$ is proportional to CNP concentration $c_{CNP}$: $A_{CNP}(\lambda) = \varepsilon_{CNP}(\lambda) l c_{CNP}$ where $\varepsilon_{CNP}(\lambda)$ are the absorptivity coefficient and $l$ the cell thickness. Given $l = 1$ cm for standard quartz SUPRASIL cell (Hellma, QS.10) and $\varepsilon_{CNP} = 25.2$ L g$^{-1}$ cm$^{-1}$ at the characteristic 288 nm CeO$_2$ peak,[48-50] the concentration of the synthesized CNPs was evaluated using the Beer-Lambert law with accuracy better than 0.1% in the range $10^{-3}$ - 10 g L$^{-1}$. For the coated particles, we ascertained that the coating did not modify the position of the peak at 288 nm nor the absorptivity and used this feature to accurately determine the CeO$_2$ concentration of the coated CNPs. In the following, all the samples studied, coated or not, will be characterized by their CeO$_2$ concentration, $c_{CNP}$.

For studies with hydrogen peroxide, the implemented protocol consisted in working at CeO$_2$ concentrations of the order of $c_{CNP} = 2 \times 10^{-2}$ g L$^{-1}$, to obtain a net absorbance at 288 nm around 0.50. Different concentrations of H$_2$O$_2$ spanning six orders of magnitude (from 0.002 mM to 2000 mM) were added to the CNP dispersions and their absorption spectra $A_{CNP+H_2O_2}(\lambda)$ were recorded. All spectra were treated by *i)* subtracting that of the solvent (H$_2$O for $A_{CNP}(\lambda)$ and mixtures of H$_2$O/H$_2$O$_2$ for $A_{CNP+H_2O_2}(\lambda)$ and *ii)* normalizing the absorbance by $A_{CNP}(\lambda = 288 \, nm)$. This data processing allowed us to compare spectra that were obtained at slightly different concentrations. Finally, the absorbance excess, defined as the difference $[A_{CNP+H_2O_2}(\lambda) - A_{CNP}(\lambda)]/A_{CNP}(\lambda = 288 \, nm)$ was calculated for each H$_2$O$_2$ concentration and plotted as a function of the wavelength. In the sequel of the paper, normalized absorbances will be written with a tilde on the $A$, such as $\tilde{A}_{CNP}(\lambda)$ or $\tilde{A}_{CNP+H_2O_2}(\lambda)$.

## II.6 – Transmission Electron Microscopy (TEM)

Micrographs were taken with a Tecnai 12 TEM operating at 120 kV equipped with a 4-k camera OneView and the GMS3 software (Gatan). A drop of the CNP dispersion was deposited on ultrathin carbon type-A 400 mesh copper grids (Ted Pella, Inc.). Micrographs were analyzed using ImageJ software for 250 particles. The particle size distribution was adjusted using a log-normal function of the form: $p(d, D, s) = 1/\sqrt{2\pi}\beta(s)d \, exp(-ln^2(d/D)/2\beta(s)^2)$, where $\beta(s)$ is related to the size dispersity $s$ through the relationship $\beta(s) = \sqrt{\ln(1+s^2)}$.[51] $s$ is defined as the ratio between the standard deviation and the average diameter. For $\beta < 0.4$, one has $\beta \cong s$.[52]

## II.7 – Fourier-Transform Infrared spectroscopy (FTIR) and Thermogravimetric analysis (TGA)

Fourier-Transform Infrared spectroscopy (FTIR) was conducted on a Spectrum65 FTIR instrument (PerkinElmer, Waltham, Massachusetts, USA) to confirm the polymer coatings onto the CNPs. The powder sample was placed on the diamond of the attenuated total reflection unit and spectra were recorded in the range $515 - 4000$ cm$^{-1}$ with 10 scans and a resolution of 4 cm$^{-1}$. Thermogravimetric analysis (TGA) was carried out on the TA Instruments SDT Q600 to determine the number of polymers bound to the cerium oxide nanoparticles. The experiments were made on powder samples of CNPs coated with PEG$_{ik}$-Ph in a N$_2$ atmosphere at a heating rate of 10 °C min$^{-1}$. Both FTIR and TGA experiments were carried out in the Phenix facilities, Sorbonne University (Paris).[53]

**II.8 - X-ray Photoelectron Spectroscopy (XPS)**

XPS analysis was performed using an Omicron Argus X-ray photoelectron spectrometer, equipped with a monochromated AlKα radiation source ($h\nu$ = 1486.6 eV) and a 280 W electron beam power. Experiments were performed on powdered samples, obtained either by heating (60° C for one day) and evaporation of solvent from the dispersion or by freeze-drying. In this way, we were able to check if heating modifies the surface defects associated with the oxygen vacancies and surface Ce(III). The emission of photoelectrons from the sample was analyzed at a takeoff angle of 45° under ultra-high vacuum conditions (≤ 10$^{-9}$ mbar). Spectra were carried out with a 100 eV pass energy for the survey scan and 20 eV pass energy for the C 1s, O 1s and Ce 3d regions. Binding energies were calibrated against the C 1s (C-C) binding energy at 284.8 eV and element peak intensities were corrected by Scofield factors. The peak areas were determined after subtraction of a Shirley background. The spectra were fitted using KolXPD software (kolibrik.net, s.r.o, Czech Republic) and applying a Gaussian/Lorentzian ratio G/L equal to 70/30. The measured core-level spectra of Ce 3d were fitted with five doublets corresponding to the Ce(III) and Ce(IV) states to evaluate the Ce oxidation state.[54] There are 3 peaks associated with Ce (IV) ions and 2 peaks associated with Ce(III) ions and each of them split in two doublets, Ce3d 5/2 (Vi ) and Ce3d 3/2 (Ui ) states, presenting a constant separation of ~ 18.5 eV. The three doublets corresponding to Ce (IV) are U''' (916.7 eV)/V''' (898.4 eV), U (901.0 eV)/V (882.5 eV), U'' (907.3 eV)/V'' (888.8 eV) and the two doublets corresponding to Ce(III) are U' (903.5 eV)/V' (884.9 eV), Uo (898.8 eV)/Vo (880.3 eV).

# III − Results and discussion

## III.1 − Structural characterization of cerium oxide nanoparticles

### III.1.1 − Transmission Electron Microscopy (TEM)

Figs. 2a-d show representative TEM micrographs of samples obtained under different conditions. For the first three, the monophosphonic acid PEG polymer PEG$_{2k}$-Ph was used as a coating agent. Fig. 2a and Fig. 2b display results at cerium-to-phosphonic-acid stoichiometry 1:1 and 1:0.1 whereas Fig. 2c was obtained at 1:1 stoichiometry, with a 4-fold lower cerium salt and polymer content, 8.8 μmol instead of 35 μmol. The CNPs featured in Fig. 2d were obtained under the same scheme using the statistical copolymer (MPEG$_{2k}$)$_6$-*co*-MPh$_5$. The size distributions displayed in the lower panels were adjusted using log-normal functions (continuous lines), leading to median diameters of 3.4, 5.1, 3.9 and 3.6 nm respectively. TEM data shows the critical role of cerium-to-phosphonic-acid stoichiometry and synthesis concentration in obtaining well-dispersed nanometric particles. When the polymer amount was decreased or when both the cerium salt and polymers were diluted, nanoparticles still formed but aggregated during synthesis to form pearl-necklace chains, and eventually anisotropic structures of several tens of nm long (Fig. 2b). The data also shows that concerning CNP cores there are no significant differences between synthesis with PEG$_{2k}$-Ph polymers and (MPEG$_{2k}$)$_6$-*co*-MPh$_5$ copolymers.

In a second series of syntheses, we studied the effect of PEG molecular weight on CNP particle size, while keeping the cerium-to-phosphonic-acid stoichiometry and polymer and cerium content of 35 μmol. The nanoparticles imaged by TEM together with the size distributions are displayed in Supporting Information S4, demonstrating the formation of cerium oxide cores for the three coatings PEG$_{1k}$-Ph, PEG$_{2k}$-Ph and PEG$_{5k}$-Ph. The median sizes of the distributions here are 3.0, 3.4 and 3.3 nm, respectively, showing no clear dependence on PEG molecular weight. The size $D_{TEM}$ and dispersity $s_{TEM}$ retrieved from the TEM experiments are shown in Table II. In the following, the cerium oxide dispersions obtained at 1:1 cerium-to-phosphonic-acid stoichiometry will be named CNPX@coating where X denotes the median TEM sizes.

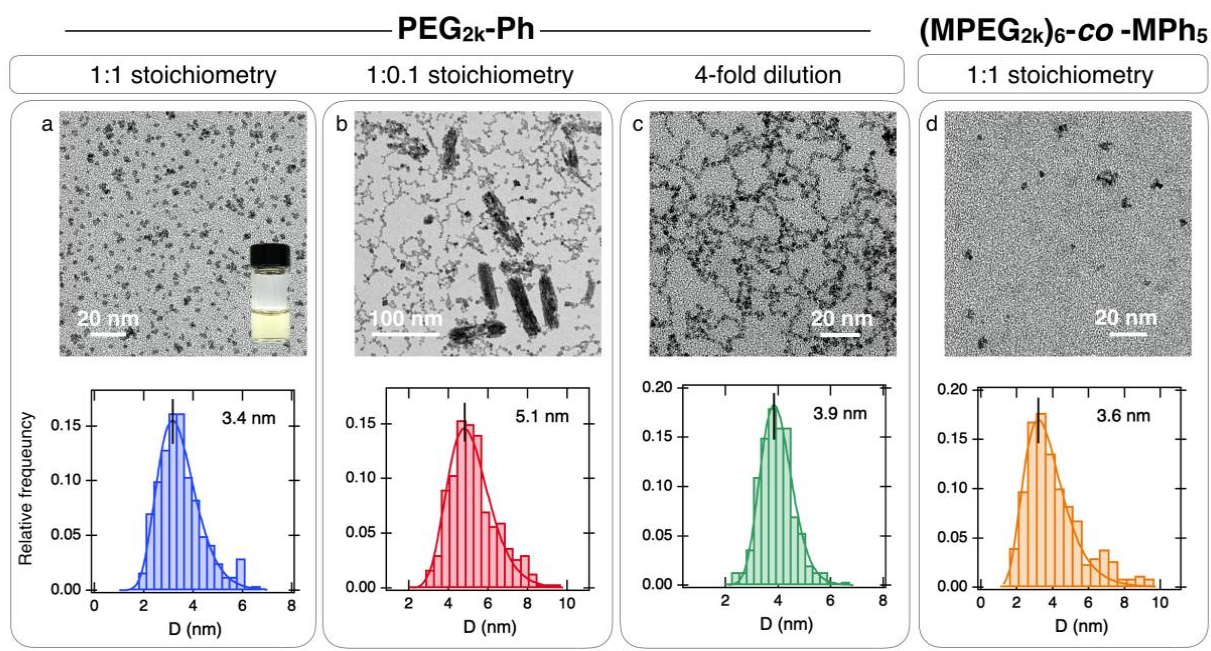

**Figure 2** : *Transmission electron microscopy images (upper panels) and corresponding size distributions (lower panels, n = 250) for (a-c) synthesis performed using PEG$_{2k}$-Ph as coating at different conditions of stoichiometry and concentration and (d) synthesis performed using statistical copolymer (MPEG$_{2k}$)$_6$-co-MPh$_5$. The continuous lines in the lower panels are the results of best-fit calculations using a log-normal function.*

### III.1.2 – Wide- and Small-Angle X-ray Scattering (SAXS, WAXS)

Wide-angle X-ray scattering was performed on CNP3@PEG$_{ik}$-Ph (i = 1, 2, and 5) powder samples and the results confirmed the face-centered cubic fluorite-like structure of the nanocrystals. From the Rietveld analysis, the lattice constants were derived and found identical for the 3 samples, at 0.5426 nm.[52] The crystallite sizes were also inferred from the X-ray spectra and found to agree with those from TEM, at 2.20 nm, 3.70 nm and 2.32 nm respectively (Table II and Supporting information S5). Fig. 3a shows data collected by powder SAXS on three batches, the CNP8 sample, and two PEGylated cerium oxide nanoparticles, CNP3@PEG$_{2k}$-Ph and CNP4@MPEG$_{2k}$-co-MPh copolymer. CNP3@PEG$_{2k}$-Ph was used to represent the set of monophosphonic PEG-coated CNPs. In SAXS, the scattering contrast arises from the electron density of the scattering species, and as the electron densities are higher for CeO$_2$ than for the coating materials, it can be assumed that the CNP cores contribute predominantly to the intensity.[27] The X-Ray scattered intensity in Fig. 3a exhibits a decrease above 1 nm$^{-1}$ for the three samples, which is consistent with a Guinier behavior, in agreement with the expression $I_{SAXS}(q) = I_0 exp(-q^2 R_g^2/3)$. In the previous equation, $q$ is the wave-vector, $I_0$ the intensity extrapolated at $q \to 0$ and $R_g$ the radius of gyration of the particles. The Guinier behavior is valid for $qR_g \ll 1$. Plotted in semi-logarithmic scale *versus* $q^2$ in Figs. 3b-d, the intensities show an exponential decay in agreement with the Guinier law. From the slope, the gyration radius $R_g$ and the effective particle diameter $D_{SAXS} = 2\sqrt{5/3}R_g$ is determined. It is found that for CNP3@PEG$_{2k}$-Ph and CNP4@MPEG$_{2k}$-*co*-MPh $D_{SAXS}$ = 4.0 and 4.7 respectively, in good accord with the TEM and WAXS data for the PEGylated CNPs (Table II). In contrast, for CNP8 results provide $D_{TEM}$ = 7.8 nm and $D_{SAXS}$ = 1.4 nm. This discrepancy can be explained by the structure of the agglomerated particles with 2 to 5 CeO$_2$ crystallites, which is in the order of 2 nm.[23,52]

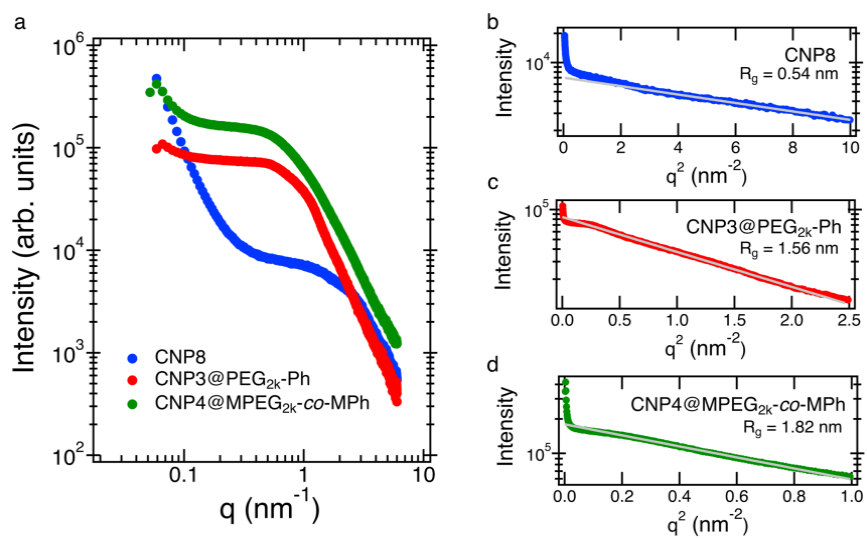

**Figure 3** : **a)** *X-Ray scattering intensity as a function of the wave vector q for CNP8, CNP3@PEG$_{2k}$-Ph and CNP4@MPEG$_{2k}$-co-MPh nanoparticles.* **b,c,d)** *Guinier representation ($I(q)$ versus $q^2$) showing the exponential decrease of the intensity in the range $qR_g \ll 1$, where $R_g$ is the radius of gyration of the particles.*

| Cerium oxide nanoparticles (CNPs) | $D_{TEM}$ (nm) | $s_{TEM}$ | $D_{WAXS}$ (nm) | $D_{SAXS}$ (nm) |
|---|---|---|---|---|
| **CNP8** | 7.8 | 0.15 | 2.56 | 1.4 |
| **CNP3@PEG$_{1k}$-Ph** | 3.0 | 0.14 | 2.20 | n.d. |
| **CNP3@PEG$_{2k}$-Ph** | 3.4 | 0.25 | 3.70 | 4.02 |
| **CNP3@PEG$_{5k}$-Ph** | 3.3 | 0.18 | 2.32 | n.d. |
| **CNP4@MPEG$_{2k}$-MPh** | 3.6 | 0.33 | n.d. | 4.71 |

**Table II** : *Diameters of the cerium oxide cores obtained by transmission electron microscopy ($D_{TEM}$) and X-ray scattering ($D_{WAXS}$, $D_{SAXS}$). $s_{TEM}$ is the size dispersity index for particles,[51] and it is defined as the ratio between the standard deviation and the mean diameter of the size distribution obtained in TEM. n.d. stands for not determined.*

III.1.3 – *Dynamic Light Scattering (DLS) and electrophoretic mobility*

DLS experiments were carried out to measure the hydrodynamic diameter of the dispersed nanoparticles. The values listed in Table III give $D_H$-values of 14 ± 2 nm, 18 ± 2 nm and 110 ± 10 nm for CNP3@PEG$_{ik}$-Ph with i = 1, 2 and 5, respectively. The value of 110 nm for CNP3@PEG$_{5k}$-Ph suggests that during synthesis the CNP have partially aggregated, a result that could not be seen from the TEM images. For CNPs coated with the copolymer (MPEG$_{2k}$)$_6$-co-MPh$_5$, the hydrodynamic diameter was found to be higher ($D_H$ = 40 ± 5 nm) than those of the monophosphonic acid polymer-coated particles. From the $D_H$-values, the thickness $h$ of the polymer layer surrounding the CNP cores can be determined (Table III). For this we estimated the hydrodynamic diameter of the bare particles based on the TEM data ($D_H^{bare} = \left(\overline{D^8}/\overline{D^6}\right)^{1/2}$),[55] and calculated $h = (D_H - D_H^{bare})/2$. For PEG$_{1k}$-Ph and PEG$_{2k}$-Ph, $h$ is 5.2 and 6.4 nm respectively, in good agreement with the thicknesses of the same polymers deposited on planar[56] or curved[39] substrates. These values are consistent with a stretching of the polymers tethered at the CeO$_2$-solvent interface of 50-80%.[36] For CNPs coated with t he copolymer (MPEG$_{2k}$)$_6$-co-MPh$_5$ of molecular weight 20.3 kDa, the coating layer value of 16 nm is higher than that of fully stretched PEG$_{2k}$ chains (12.7 nm),[32] suggesting that phosphonic acid groups are not all anchored to the surface, some of them being dangling in the solvent. For partially aggregated CNP3@PEG$_{5k}$-Ph particles, it was not possible to retrieve $h$, as the diameter of the CNP aggregates could not be pinpointed from TEM micrographs. In Table III we have also added data from ζ-potential which show slightly negative values comprised between ζ = -0.3 and - 8.1 mV. The intensities *versus* ζ-potential obtained for the 4 coated CNPs are displayed in Supporting Information S6. Finally, CNPs powders obtained by freeze-drying were examined by Fourier-transform infrared spectroscopy (FTIR, Supporting Information S7) and thermogravimetric analysis (TGA, Supporting Information S8) to confirm polymer coating of CNPs and to determine the percentage of organic matter associated with the particles. FTIR spectra of PEG$_{ik}$-Ph coated NPCs reveal absorption bands at 2860 - 2870 cm$^{-1}$ due to aliphatic C-H stretching. The aliphatic C-H stretching at 1500 - 1300 cm$^{-1}$ is due to C-H bending vibrations. The bands due to the C-O/P-O stretching mode were observed between 1300 and 1000 cm$^{-1}$ and originate from C-O vibrations, C-O-C stretching and C-O-H bending vibrations of PEG$_{ik}$-Ph. The peaks between 960 and 800 cm$^{-1}$ are related to out-of-plane C-H vibrations (wobbling and swinging) of the PEG$_{ik}$-Ph chains, while the absorption peak around 720 cm$^{-1}$ can be attributed to P-C stretching. After the coating of the polymers on the CNPs, the FTIR spectra confirm the successful incorporation of these polymers on the CNP surface. As for the TGA, the percentage of organic matter indicated values between 45 and 83% depending on the polymers (Supporting Information S8). The numbers of polymers per CNP are found between 50 and 100 (Table III), corresponding to polymers densities around 1 nm$^{-2}$. Such densities are common for coated particles[31,57-58] and slightly larger than those obtained from the two-step post-synthesis process, for which the polymers are grafted after nanoparticle synthesis.[20,22] In conclusion of this part, we find that under the synthesis conditions used, PEG$_{1k}$-Ph and PEG$_{2k}$-Ph polymers provide a coating whose conformation can be described in terms of a stretched polymer brush.[59]

| Cerium oxide nanoparticle (CNP) | $D_H$ (nm) | $h$ (nm) | ζ (mV) | Number of polymers per CNP |
|---|---|---|---|---|
| **CNP3@PEG$_{1k}$-Ph** | 14 ± 1 | 5.2 | -8.1 | 58 |
| **CNP3@PEG$_{2k}$-Ph** | 18 ± 2 | 6.4 | -2.7 | 57 |
| **CNP3@PEG$_{5k}$-Ph** | 110 ± 10 | n.d. | -0.3 | 107 |

| CNP4@MPEG$_{2k}$-MPh | 40 ± 5 | 16.2 | -5.2 | 57 |
| --- | --- | --- | --- | --- |

***Table III*** : *Hydrodynamic diameters ($D_H$) and thickness (h) of the polymer coating layer of cerium oxide particles synthesized in this work. The zeta potential ζ was derived from electrophoretic mobility measurements (Supporting Information S6) whereas the number of polymers per particles was acquired from thermogravimetric analysis (Supporting Information S8). n.d. stands for not determined.*

### III.2 – X-ray photoelectron spectroscopy (XPS)

A feature commonly associated with the antioxidant properties of CeO$_2$ is the fraction $f_{Ce(III)}$ of surface cerium atoms in oxidation state Ce(III).[60] For the determination of $f_{Ce(III)}$, X-ray photoelectron spectroscopy experiments were performed on powdered samples obtained by solvent evaporation (60 °C for days) or by freeze-drying of CNP dispersions. Figs 4a display XPS Ce3d spectra for CNP8 powders obtained from a 10 g L$^{-1}$ dispersion using solvent evaporation. The data was adjusted using the model proposed by Y. Lykhach *et al.*,[54] assuming three pairs of peaks associated with Ce3d electrons ejected from Ce(IV) and two for Ce(III). Also represented in the figures are the separate contributions from each oxidation state, illustrated by the shaded areas in red and blue respectively. For this sample, we found a $f_{Ce(III)}$ at 16%, in good agreement with that obtained in an earlier report.[52] The uncertainty on the Ce(III) fraction with the method of Y. Lykhach et *al.* is estimated to be ± 5%.[54] Samples obtained by freeze-dry preparation were also measured and evaluated for the CNP8 sample to be 8%. Both methods gave similar results of Ce(III) fraction with standard deviation fitting within the error estimation range.

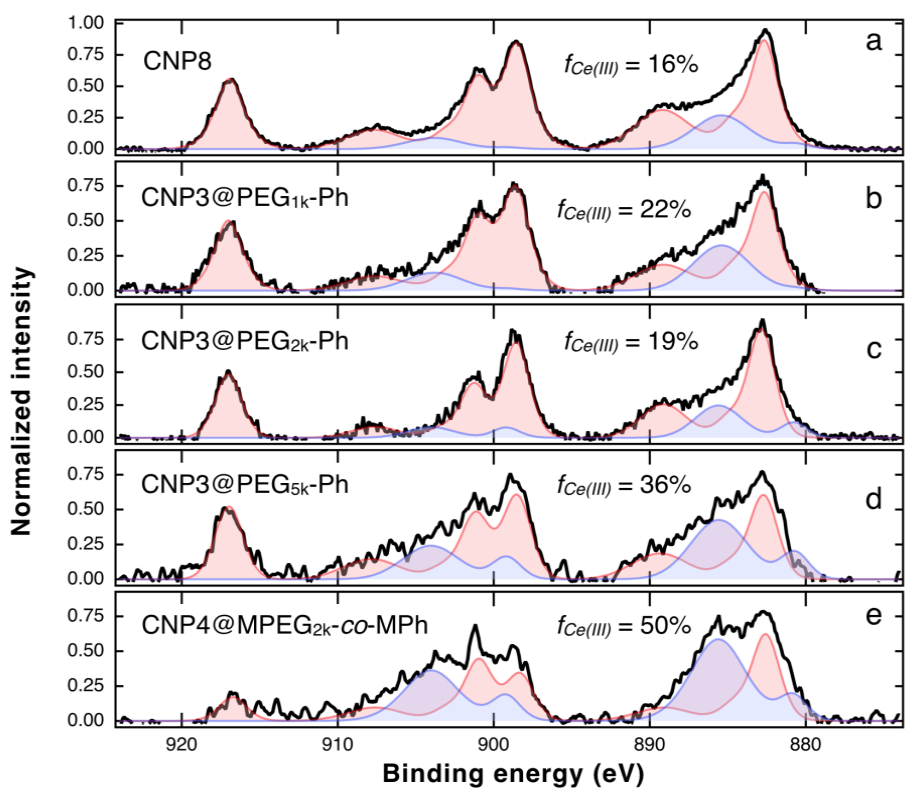

*Figure 4*: a) X-ray photoelectron spectroscopy of Ce3d core-level spectra for CNP8 powders obtained from a 10 g L$^{-1}$ dispersion. The continuous thick lines display the sum of the different peak contributions adjusted according to the model proposed by Y. Lykhach et al..[54] The continuous thin lines in red and green refer to the Ce(IV) and Ce(III) contributions respectively. The Ce(III) fractions were calculated from the integrated areas of the assigned peaks. b,c,d,e) Same as in a) for CNP3@PEG$_{1k}$, CNP3@PEG$_{2k}$, CNP3@PEG$_{5k}$ and CNP4@MPEG$_{2k}$-co-MPh.

Figs. 4b-e show the decomposed Ce3d XPS spectra for the four coated nanoparticles, CNP3@PEG$_{1k}$-Ph, CNP3@PEG$_{2k}$-Ph, CNP3@PEG$_{5k}$-Ph and CNP4@MPEG$_{2k}$-*co*-MPh, respectively. As above, the contributions of two oxidation states have been included. For CNP3@PEG$_{1k}$-Ph, CNP3@PEG$_{2k}$-Ph, CNP3@PEG$_{5k}$-Ph and CNP4@MPEG$_{2k}$-*co*-MPh, one gets $f_{Ce(III)}$ = 22%, 19%, 36% and 50% respectively. The high Ce(III)-fraction is recognizable on the graphs as it shows the prominent (U',V') doublet at energies of 903/885 eV. These results show that the one-pot wet-chemical synthesis provides more reduced CNPs, with $f_{Ce(III)}$ in the range of 19-36%. As found in the literature, large Ce(III) fractions are associated with smaller CNP core sizes, in relation to increased surface lattice deformation and oxygen vacancies.[10,61] Furthermore, polymer interaction with surface Ce ions has been reported to mediate the surface oxidation state of CeO$_2$, which largely depends on the interaction mechanisms between the polymer functional groups and the surface oxygen vacancies.[12] High $f_{Ce(III)}$-values for the synthesized CNPs represent a major result as the outcome suggests that this nanomaterial could be of interest for biomedical applications.

### III.3 – UV-Visible spectrophotometry

It is well-known that CNP dispersions change their spectroscopic properties upon H$_2$O$_2$ addition.[6,12,14,62-64] For instance, Fig. 5a shows the colorimetric alteration of an 8 g L$^{-1}$ CNP8 dispersion with increasing [H$_2$O$_2$] from 0 to 200 mM. The CNP8 dispersions tend to change from yellowish to orange and from orange to garnet as [H$_2$O$_2$] increases. Previous reports have interpreted these color changes as a red-shift, that is a shift of the absorbance spectrum to higher wavelengths.[12,14,16,52,64] The first approach for quantifying these results consisted in measuring the shift in wavelength at the absorbance of 0.3 before and after the addition of peroxide, and comparing this shift across different samples.[14] Here, we opted for the method proposed by Damatov *et al.*[42] who evaluated the absorbance excess $\tilde{A}_{exc}(\lambda) = \tilde{A}_{CNP+H_2O_2}(\lambda) - \tilde{A}_{CNP}(\lambda)$ after addition of H$_2$O$_2$ on the entire $\lambda$-range. In the previous expression, the tilde indicates that the data have been normalized with respect to $A_{CNP}(\lambda = 288\ nm)$.

As part of this work, we investigated the effect of H$_2$O$_2$ on the CNP absorption spectrum as a function of the wavelength. For this purpose, we compared CNPs with different sizes and coatings, *i.e.* CNP8 and the PEGylated CNPs. In these assays, the CeO$_2$ concentration was set at $c_{CNP}$ = 2×10$^{-2}$ g L$^{-1}$ and that of hydrogen peroxide was increased from 2×10$^{-3}$ mM to 2×10$^3$ mM. The upper panel in Fig. 5b shows the normalized absorbance excess $\tilde{A}_{exc}(\lambda)$ for CNP8 as a function of the H$_2$O$_2$ concentration. With increasing hydrogen peroxide concentration, an increase in absorbance is observed in the 300-500 nm region, as indicated by an upward arrow. At lower wavelengths, the spectra are well overlapped, except those at 200 mM and 2000 mM which do not display the 288 nm peak. We ascribe the missing 288 nm peaks to the increase in the absorbance of H$_2$O$_2$ in the UV region. This increase induces a spectral bias in this range, especially after the subtraction of the solvent. The lower panel of Fig. 5b shows the absorbance excess $\tilde{A}_{exc}(\lambda)$ in the same spectral range as Fig. 5a. The data discloses the existence of an absorbance peak centered around 365 nm, which amplitude increases with $H_2O_2$ concentration (Supporting Information S9). The normalization of the 365 nm peak by its maximum shows a good superposition of the data, suggesting that this band is associated with a single mechanism (Supporting Information S10).

The same protocol and data treatment were applied to the PEGylated CNPs, CNP3@PEG$_{1k}$-Ph, CNP3@PEG$_{2k}$-Ph, CNP3@PEG$_{5k}$-Ph and CNP4@MPEG$_{2k}$-*co*-MPh, revealing features similar to CNP8. In Figs. 5c and 5d, an increase in absorbance in the 300-500 nm region is observed upon increasing [$H_2O_2$] resulting in the absorption band centered again around 365 nm. Once normalized, these peaks are superimposable to each other. In Supporting Information S10, we demonstrate that the 365 nm band has the same shape and width characteristics independent of the type of investigated particles. In conclusion of this section, for bare and PEGylated CNPs, we find identical features for the spectrophotometric properties induced by $H_2O_2$, namely an additional absorbance band centered at $\lambda$ = 365 nm, attributed to the formation of stable surface Ce-O$_2^{2-}$ peroxo-complexes.[42-47,52,65]

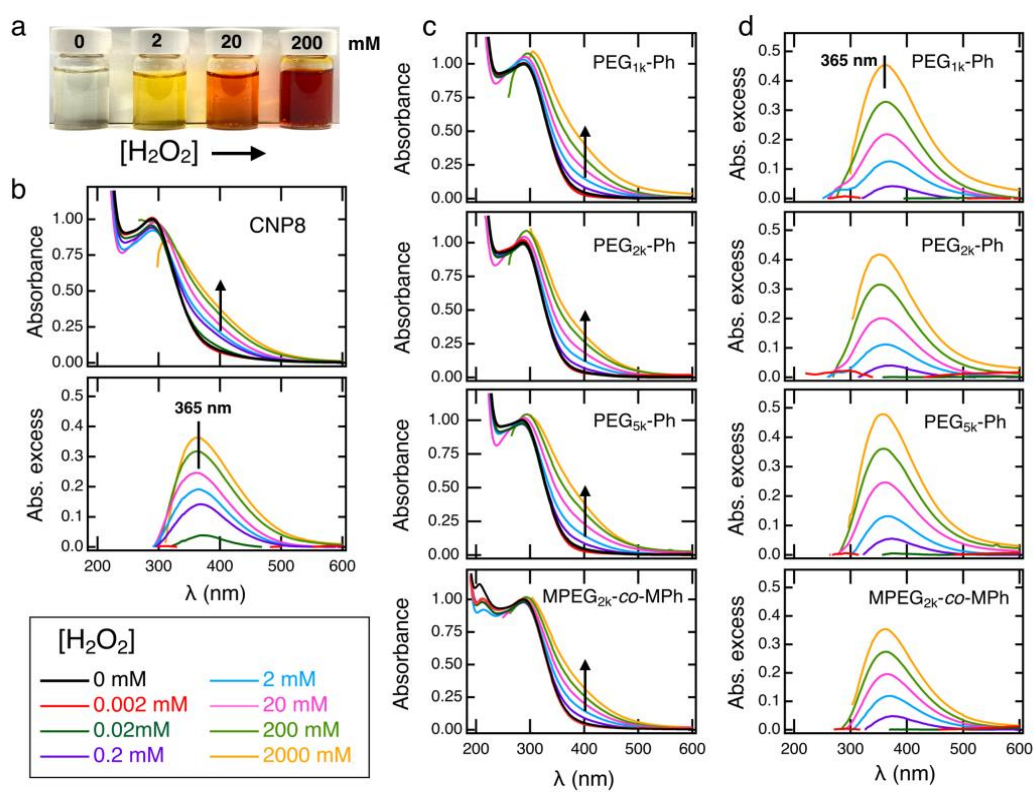

**Figure 5** : *a) Cerium oxide nanoparticles (CNP8) at the concentration of 8 g L$^{-1}$ with increasing addition of $H_2O_2$. From left to right, [$H_2O_2$] = 0, 2, 20 and 200 mM. b) Upper panel: Normalized absorption spectra $\tilde{A}_{CNP+H_2O_2}(\lambda)$ obtained for CNP8 at the cerium oxide concentration of $c_{CNP}$ = 2×10$^{-2}$ g L$^{-1}$ and hydrogen peroxide concentration ranging from 2×10$^{-3}$ mM to 2×10$^3$ mM. Lower panel: Absorbance excess $\tilde{A}_{CNP+H_2O_2}(\lambda) - \tilde{A}_{CNP}(\lambda)$ as a function of the wavelength at the same [$H_2O_2$] concentrations. c) Same as in Fig. 5b upper panel for CNP3@PEG$_{1k}$-Ph, CNP3@PEG$_{2k}$-Ph, CNP3@PEG$_{5k}$-Ph and CNP4@MPEG$_{2k}$-co-MPh nanoparticles. d) Same as in Fig. 5b lower panel for CNP3@PEG$_{1k}$-Ph, CNP3@PEG$_{2k}$-Ph, CNP3@PEG$_{5k}$-Ph and CNP4@MPEG$_{2k}$-co-MPh nanoparticles.*

Fig. 6a-e depict the variation of the absorbance excess $\tilde{A}_{exc}([H_2O_2])$ at $\lambda$ = 365 nm as a function of $H_2O_2$ concentration for the 5 samples studied previously, CNP8, CNP3@PEG$_{ik}$-Ph (i = 1, 2, 5) and CNP4@MPEG$_{2k}$-*co*-MPh. We observe that $\tilde{A}_{exc}([H_2O_2])$ remains close to zero until the concentration of $[H_2O_2]$ = $10^{-2}$-$10^{-1}$ mM, then increases rapidly until reaching a plateau at the highest concentration, *i.e.* above 100 mM. This behavior is similar to that found by Damatov and Mayer for dispersions of $CeO_2$ in non-polar organic solvents.[42] It should be noted that all three samples coated with monophosphonic acid PEG chains behave similarly, whereas the CNP4@MPEG$_{2k}$-*co*-MPh sample lies slightly lower. The $\tilde{A}_{exc}([H_2O_2])$ behavior was adjusted with the Hill adsorption isotherm:[66]

$$\tilde{A}_{exc}([H_2O_2]) = \tilde{A}_{max}\frac{[H_2O_2]^\alpha}{c_0^\alpha + [H_2O_2]^\alpha} \qquad (Eq. 1)$$

where $\tilde{A}_{max}$ is the maximum value of absorbance excess, $c_0$ the $H_2O_2$ concentration at which $\tilde{A}_{exc}([H_2O_2])$ is at half maximum and $\alpha$ is the Hill coefficient. For all investigated CNPs, $\alpha$ was found less than 1, varying between 0.37 and 0.42 (Fig. 6f), suggesting a negative cooperativity effect for the $H_2O_2$ binding to the surface.[67-68] Despite the polymer coating, the 3 nm core CNPs exhibits higher maximum absorbance compared to CNP8 with a core size of 7.8 nm, a result that could be attributed to the higher number of exposed Ce(III) ions to the surface, as evidence by XPS.[52] On the other hand, the value of $c_0$ for CNP8 is one order of magnitude lower than that of PEGylated CNPs. This suggests that the polymer coatings on CNP3@PEG$_{1k}$-Ph, CNP3@PEG$_{2k}$-Ph, CNP3@PEG$_{5k}$-Ph and CNP4@MPEG$_{2k}$-*co*-MPh affect the diffusion of $H_2O_2$ to the surface.

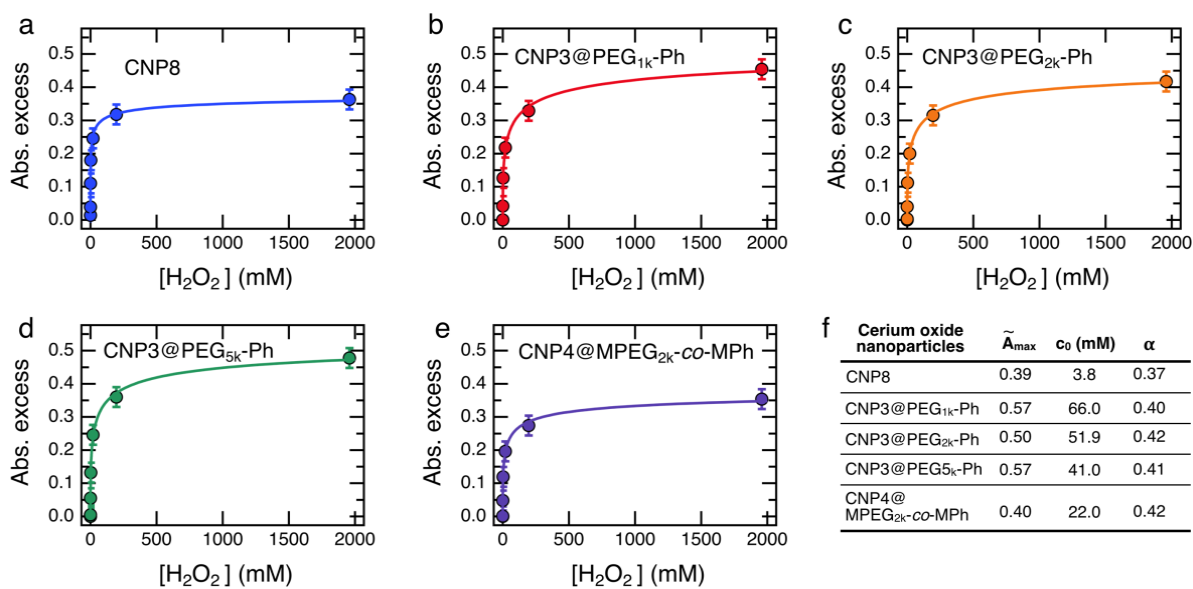

| Cerium oxide nanoparticles | $\tilde{A}_{max}$ | $c_0$ (mM) | $\alpha$ |
| --- | --- | --- | --- |
| CNP8 | 0.39 | 3.8 | 0.37 |
| CNP3@PEG$_{1k}$-Ph | 0.57 | 66.0 | 0.40 |
| CNP3@PEG$_{2k}$-Ph | 0.50 | 51.9 | 0.42 |
| CNP3@PEG$_{5k}$-Ph | 0.57 | 41.0 | 0.41 |
| CNP4@ MPEG$_{2k}$-*co*-MPh | 0.40 | 22.0 | 0.42 |

**Figure 6** : *a-e) Absorbance excess $\tilde{A}_{CNP+H_2O_2} - \tilde{A}_{CNP}$ at the wavelength $\lambda$ = 365 nm after addition of increasing amounts of $H_2O_2$ and 1 hour equilibration time for CNP8, CNP3@PEG$_{1k}$-Ph, CNP3@PEG$_{2k}$-Ph, CNP3@PEG$_{5k}$-Ph and CNP4@MPEG$_{2k}$-co-MPh respectively. The continuous line was obtained from the Hill equation (Eq. 1). f) Hill parameters used to fit the absorbance excess as a function of the $H_2O_2$ concentration.*

### III.4 – Stability in cell culture media

The biomedical application of CNPs requires robust colloidal stability of the nanoparticles in the biologically relevant environment. What the cell "sees" after nanoparticles enter the human biological fluid can be significantly different from their physico-chemical properties from the bench synthesis.[12] For this reason, we further evaluated the colloidal stability of these polymer-coated CNPs using representative cell culture media. It was achieved by examining the state of the CNP dispersions as a function of time, over a period from a few seconds to several weeks. To this aim, static and dynamic light scattering was implemented. The cell culture medium used was Dulbecco's Modified Eagle Medium (DMEM) supplemented with 10 vol. % fetal bovine serum (FBS) and antibiotics (penicillin-streptomycin). DMEM is representative of cell biology experiments that deal with nanoparticles for toxicology or biomedical purposes. Based on recently established protocols,[21] a dispersion is considered stable if the scattered intensity $I_S(t)$ and the hydrodynamic diameter $D_H(t)$ obtained are both stationary in time, and if $D_H(t)$ is identical to its value in DI water. As for the protocol, 100 µL of a $c$ = 20 g L$^{-1}$ concentrated dispersion was poured and homogenized rapidly in 1 mL of the solvent to be studied, and $I_S(t)$ and $D_H(t)$ are simultaneously measured over 2 hours and later at day 1, 2, 3, 5, 6, 10, 14, 20 and 26. This procedure ensures that the scattering intensity of the CNPs is much higher than that of the proteins in the culture medium, and that the measured $D_H$s are indeed those of the nanoparticles.

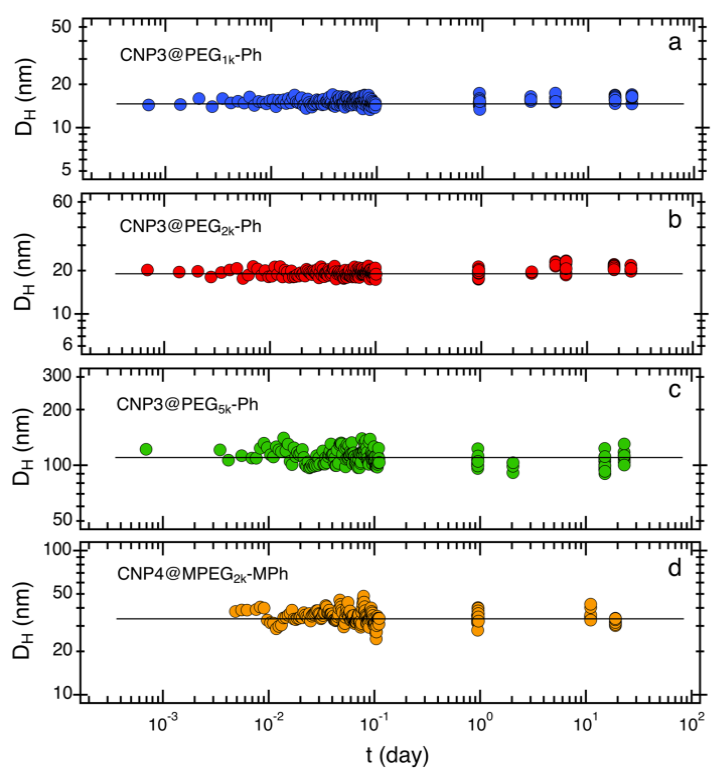

***Figure 7*** *: Time dependence of the hydrodynamic diameter of a) CNP3@PEG$_{1k}$-Ph, b) CNP3@PEG$_{2k}$-Ph, c) CNP3@PEG$_{5k}$-Ph, and d) CNP4@MPEG$_{2k}$-co-MPh in a cell culture medium (Dulbecco's Modified Eagle Medium with 10 vol. % FBS and antibiotics) over a period of 26 days.*

*The horizontal straight lines indicate time average values, which are in agreement with those measured in DI-water (Table II).*

Fig. 7 shows the time dependence of CNP3@PEG$_{ik}$-Ph (i = 1,2, 5) and CNP4@MPEG$_{2k}$-*co*-MPh diameters for particles dispersed in complete DMEM. We find that for the 4 samples, the $D_H$-behavior as a function of the time meets the criteria defined previously: the hydrodynamic diameter is constant over time and the values are those obtained after synthesis (Table II). For comparison, we show in the Supporting Information how the CNP8 sample without polymer stabilization behaves when brought into contact with the cell medium (Movie #1). The movie illustrates that upon contact with the solvent, the dispersion instantly destabilizes and the particles aggregate into micron-sized particles. This study concludes that the particles synthesized in the presence of phosphonic acid PEG polymers are colloidally stable and devoid of the protein corona.

# IV – Conclusion

A convenient strategy for coating cerium oxide nanoparticles in the liquid phase is to allow hydrophilic polymer terminated by a single anchoring group to adsorb spontaneously to the particle surface, e.g. by electrostatic interaction. From a polymer chemistry perspective, adding a binding group to the end of a PEGylated chain is a relatively simple process, and such polymers could represent an all-in-one solution for coating a wide range of metal oxide nanoparticles. Our previous work has shown that although monophosphonic acid PEG polymers adsorb well on CNP surfaces during post-grafting synthesis, they did not stabilize CNPs in protein-rich cell culture media, the reason being the progressive desorption of the polymer adlayer, the formation of a protein corona and finally the particle aggregation[20] In this work, we address this problem following an alternative coating route, namely through the one-step wet-chemical synthesis where the polymers are introduced along with cerium precursors during the synthesis. Combining different techniques such as transmission electron microscopy, X-ray, and light scattering, the synthesized CNP3@PEG$_{ik}$-Ph (i = 1,2,5) were found to exhibit a core-shell structure, where the CeO$_2$ cores are about 3 nm in size and the shell consists of PEG polymers in a brush configuration. The most promising results were observed with monophosphonic acid PEG polymers of molecular weight 1 and 2 kDa while keeping a 1:1 stoichiometry between Ce(III) ions and phosphonic acid moieties. We found that stoichiometry and initial concentrations also played an important role in obtaining well-dispersed particles. CNP3@PEG$_{5k}$-Ph showed also 3 nm cores, but partial particle agglomeration during synthesis, with sizes reaching around 100 nm. We also performed a one-step synthesis with the (MPEG$_{2k}$)$_6$-co-MPh$_5$ copolymer for comparison and obtained results that were globally equivalent to those of single phosphonic acid PEG polymers. The combination of different characterization techniques, such as X-ray photoelectron spectroscopy for the quantification of the Ce(III) fraction, UV-vis spectrophotometry, and colloidal stability assays by light scattering techniques, showed that CNP3@PEG$_{1k}$-Ph and CNP3@PEG$_{2k}$-Ph are of potential interest for biomedical applications, thanks to their elevated Ce(III) fractions and improved stability in cell culture media. This increased stability could also result from a higher density of PEG polymers on the surface compared to the two-step post-synthesis process, for which polymers are grafted after nanoparticle synthesis. We also investigated the colorimetric changes of CNPs upon $H_2O_2$ addition by UV-Vis spectrophotometry, and found that the absorbance spectra contained a band at 365 nm in all coated and uncoated samples. This band was attributed to the formation of stable $Ce-O_2^{2-}$ peroxo-complexes[42-45,47] and could provide quantitative information on CNP catalytic properties. A systematic study of CNP absorption spectra as a function of $H_2O_2$ molar concentration showed an increase in absorbance at 365 nm in agreement with the Hill adsorption isotherm.[66] From this analysis, a comparison between bare and coated samples suggest that the polymer coating affects the diffusion of $H_2O_2$ to the surface. In conclusion, we find that with a one-step wet-chemical synthesis and a convenient choice of phosphonic acid-terminated polymers, it is possible to synthesize cerium oxide nanoparticles as antioxidants with potential beneficial properties for nanomedicine applications.

## Supporting Information

Transmission electron microscopy of 7.8 nm cerium oxide nanoparticles (S1); size exclusion chromatography on PEG$_{ik}$-Ph precursors (S2); $^1$H and $^{31}$P NMR spectra from phosphonic acid PEG polymers, copolymers and phosphonate ester precursor (S3); transmission electron microscopy of CNP3@PEG$_{ik}$-Ph (i = 1, 2 and 5) cerium oxide nanoparticles (S4); wide-angle X-ray scattering (WAXS) performed on CNP3@PEG$_{ik}$-Ph nanoparticles (S5); electrophoretic mobility of cerium coated oxide nanoparticles (S6); Fourier-transform infrared spectroscopy on CNP3@PEG$_{ik}$-Ph (S7); thermogravimetric analysis (S8); peroxo-complex peak position as retrieved from the UV-Vis spectrometry (S9); normalized excess absorbance induced by the addition of $H_2O_2$ (S10).

## Acknowledgments

We thank Victor Baldim and Jean-Yves Piquemal for fruitful discussions. Delphine Talbot from the PHENIX laboratory (Sorbonne University) is also acknowledged for the thermogravimetric analyses. ANR (Agence Nationale de la Recherche) and CGI (Commissariat à l'Investissement d'Avenir) are gratefully acknowledged for their financial support of this work through Labex SEAM (Science and Engineering for Advanced Materials and devices) ANR-10-LABX-0096 et ANR-18-IDEX-0001. We acknowledge the ImagoSeine facility (Jacques Monod Institute, Paris, France), and the France BioImaging infrastructure supported by the French National Research Agency (ANR-10-INBS-04, « Investments for the future »). This research was supported in part by the Agence Nationale de la Recherche under the contract ANR-13-BS08-0015 (PANORAMA), ANR-12-CHEX-0011 (PULMONANO), ANR-15-CE18-0024-01 (ICONS), ANR-17-CE09-0017 (AlveolusMimics) and by Solvay.

## TOC Image

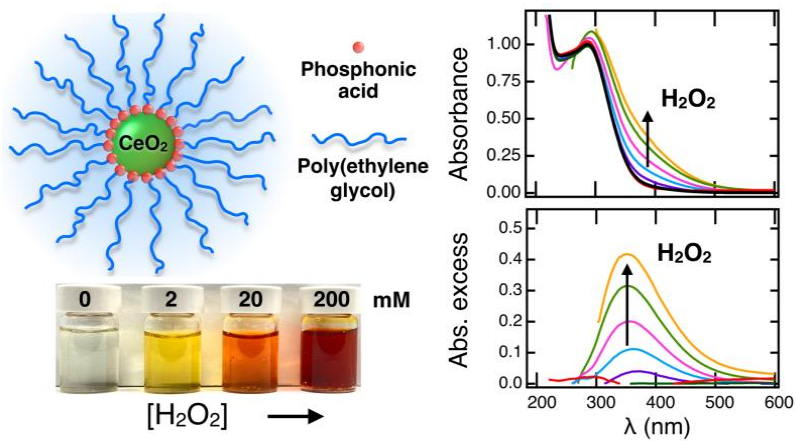